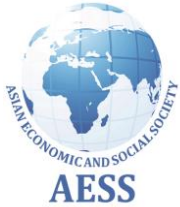

**International Journal of Asian Social Science**

**Special Issue:** International Conference on Teaching and Learning in Education, 2013

journal homepage: http://www.aessweb.com/journal-detail.php?id=5007

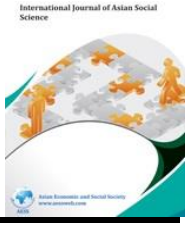

# A NOVEL INTERACTIVE OBE APPROACH IN SCM PEDAGOGY USING BEER GAME SIMULATION THEORY

**S.E.S.Bariran**

*Centre for Advanced Mechatronics and Robotics,College of Engineering,UNITEN ,Kajang,Selangor,Malaysia*

**K.S.M.Sahari**

*Centre for Advanced Mechatronics and Robotics,College of Engineering,UNITEN ,Kajang,Selangor,Malaysia*

**B.Yunus**

*Strategic and Corporate planning Centre,UNITEN ,Kajang, Selangor,Malaysia*

## ABSTRACT

*The primary challenge in SCM pedagogy is the learners' interaction with the dynamic nature of supply chain transactions. Once achieved, it is also required to evaluate learners' learning experience based on their performance. In this paper, a combination of outcome-based education (OBE) and simulation-based education is proposed focusing on beer game theory. The analysis is based on 336 runs of beer game simulation within a target group of 56 participants divided into 14 subgroups (SG1-SG14).The purpose of the study is mainly to investigate the effect of mutual interactions on students' learning process using supply chain total cost and ordering fluctuations as critical measurement criteria.*

## 1. INTRODUCTION

Outcome-based education (OBE) is known as an alternative pedagogy for the traditional teacher-centered education. It is primarily based on shifting the focus from inputs or students' available resources to empirical measurement of student's performance or outputs. Performance-based learning is another terminology that is sometimes used to refer to the same conceptual framework in today's modern education era. An effective outcome-based approach is usually characterized by three main components: (1) an explicit and measurable learning outcome, (2) a strategy-driven process to attain the outcomes and finally (3) an explicit assessment/measurement criteria (Nicholson, 2011).

An evolutionary review identifies at least three phases for outcome-based education (Biggs & Tang, 2011). Initially, the term OBE was coined by Spady (Killen, 2000) during 80's decade. Later, during and 90's, OBE entered the period of ensuring accountability which required some performance indicators that were defined as inputs and outputs. And finally at the third phase, OBE was used as a tool to enhance the teaching and learning process (Nicholson, 2011).

OBE can be applied in different fields of study especially where there is a lack of systematic approach for evaluating the final collective outputs of the working system. Supply chain management (SCM) is a good example of such systems that due to its insular distribution-based mechanism, measurement of overall performance is a matter of challenge and therefore many critical assessment criteria may simply be ignored as a result of human errors, increased complexity and unpredictable fluctuations in the supply chain. Therefore, a SCM pedagogy that is solely based on theoretical/mathematical models and concepts may even add to the complexity of learning process (Sparling, 2002).

Learners need to practically experience the interactions that occur in a real supply chain for better timely responses. In each supply chain, regardless of the size and complexity, there is a continuous mutual effect among components. In other words, a reciprocal learning process exists throughout the whole supply chain as members track and monitor their upstream and downstream followers. In traditional methods of teaching supply chain, this concept is rarely paid attention due to the teaching environment that is far distant from the atmosphere of a real supply chain (Sparling, 2002).

Beer game software is a soft skill tool that can simply be used in traditional classrooms to simulate a real supply chain in order to investigate some challenging concepts such as bullwhip effect and components mutual interactions. In this study we will use the beer game theory for two general objectives. Firstly, beer game is used in classrooms as an effective outcome-based evaluation tool for individual measurement of leaner's progress based on weekly simulation reports generated by the software. And secondly, the research will focus on the relationship between learner's interactions and overall performance based on a real case study.

A group of 56 students from MBA and engineering management background is randomly selected and divided into 2 main groups of A & B. Each group has 7 subgroups (SG1-7 for group A and SG8-14 for group B) of 4 members. Then, both groups will simulate the same supply chain scenario for a period of 24 weeks using the same SCM theory with the only difference that for subgroups of "group B" (SG8-14), interaction is prevented while subgroups of "group A" can freely interact during the simulation process. Finally, groups' performance will be judged based on their respective total cost and ordering fluctuations as two measurement criteria. It is predicted that due to the significant constructive effects of open interactions, group A will have a better overall performance as compared to group B.

## 1.1. Beer game Background and Structure

The Beer Game is a role-playing simulation developed at MIT in the 1960's to clarify the advantages of taking an integrated approach to supply chain management. The game can be played either manually which is called as "traditional Beer Game" or the "online computerized version". In fact, the computerized Beer game has been made to make it easier to play the Beer Game as well as to illustrate certain Supply Chain Management issues which cannot be demonstrated by the traditional (non-computerized) Beer Game. The Beer Game supply chain consists of four typical components: (1) retailer has to fulfill the end consumer's orders, (2) wholesaler has to fulfill the retailer's orders, (3) distributor has to fulfill the wholesaler's orders and (4) factory has to fulfill the distributor's orders (Simchi-Levi et al., 1999).

### 1.2. Simulation Procedure and Rules

The simulation runs on a weekly basis starting at week 1. During each run, any component in the supply chain tries to satisfy the demand of its immediate downstream follower. Orders which cannot be met at a certain week are recorded as backorders, and met as soon as possible. No orders will be ignored, and all orders must eventually be met. Once the order arrives, the supplier attempts to fill it with available inventory, and there is an additional two week transportation delay before the material being shipped by the supplier arrives at the customer who placed the order. There are two different kinds of cost: (a)-Inventory cost: Items in stock cost unit 0, 50 per week in holding costs and backorder cost: If an incoming order cannot be (fully) fulfilled, items are outstanding and have to be put on "backorder" to be fulfilled in the following week(s). Each item on backorder costs unit 1, 00 per week (Simchi-Levi et al., 1999).

## 2. A REVIEW OF SIMULATION TOOLS USED IN SCM PEDAGOGY

This section provides a summary of major simulation tools that have more frequently been used in SCM pedagogy such has beer game simulation, electronic data interchange (EDI) and discrete event simulation (DES).

**Table- 1.** Review of simulation tools in SCM

| Researcher(s)/ year | Brief Description / Major findings |
|---|---|
| (Machuca & Barajas, 2004) | The effect of comprehensive use of electronic data interchange (EDI) on supply chain and the bullwhip effect indicates a significant reduction in all parameters related to cost such as mean inventory and cumulative cost as well as bullwhip effect indicators such as amplification and net excess stock |
| (Wu & Katok, 2006) | - Using beer game supply chain to study the effect of learning and communication on bullwhip effect<br>- results indicate that order variability is dramatically reduced in cases that the simulation is initiated individually and then continued as a collaborative work<br>-Training is also proved to affect supply chain performance only if the knowledge is shared among partners and that lack of coordination among partners is one of the main reasons of bullwhip effect in beer distribution game supply chain |
| (Nienhaus et al.,2006) | Online beer game simulation involving a group of 400 people suggest that the interactive role of human behaviour should be added as one of the potential causes of bullwhip effect as compared to other simple agent-based approaches |
| (Paik & Bagchi,2007) | A survey on nine potential causes of bullwhip effect proves "demand forecast updating, level of echelons, and price variations" as three significant co-factors |
| (Hussain et al., 2012) | -Adopting a system dynamic methodology and a complementary *iThink*® software<br>-Suggesting a non-monotonic relationship between batch size and demand magnification<br>- Information sharing is also found to be an adding value for smaller batch size supply chains |
| (Tako & Robinson, 2012) | Introducing discrete event simulation (DES) as is widely used for modelling supply chains |

## 3. SCM METHODOLOGY USED IN SIMULATION

The simulation in is based on s-Q policy in supply chain meaning that whenever the system inventory level falls below the value of "s", an automatic order of "Q" will be placed and during simulation period (24 weeks runs), demand is based on random normal distribution with a 1 week fixed time elapse between each two orders. The inventory level is reviewed periodically at regular intervals (weekly) and an appropriate quantity will be ordered after each review. Since the order is placed after each inventory review the fixed cost of placing an order can be neglected and the quantity ordered arrives after the appropriate lead time (2weeks).To find the effective base-stock level the methodology behind the game is as what follows:

Let R as the length of the review period, L as lead time, AVG is the average weekly demand face by the player z as the value of normal distribution and STD is the standard deviation of this weekly demand. At the time the downstream component places an order this order raises the inventory position to the base stock level. This level of inventory should be enough to protect the player against shortages until the next order arrives. Since the next order arrives after a period r + L days, the current order should be enough to cover the demand during a period of r + L days. Using the aforementioned variables, average inventory level (AIL) can be calculated as in Equation (1).

$$AIL = (R * AVG * z) / 2 * STD * (R + L)^{1/2} \qquad (1)$$

## 4. SIMULATION SCENARIO

A group of 56 participants with SCM background are divided into two groups of A & B each consisting of 7 subgroups. The interactive roles (highlighted cells in Tables 2 and 3) may vary during simulation but, both groups will have the same interactive roles for their matching subgroups (i.e., SG1&SG8, SG2&SG9, etc). For all groups and subgroups, the game rules and SCM theory will be identical to maintain the homogeneity of the simulation. The only key difference is that the for group B, unlike group A, the simulation options are set in such a way that the players cannot view information such as latest total cost, back order and recent order history, related to their upstream and downstream roles except for the immediate downstream of the interactive role, like for wholesaler if interactive role is retailer. For players in group A, all supplementary information related to other interactive roles is observable that allows for a better internal interaction between supply chains members. Furthermore, group A also has the privilege that simulations are conducted consequently rather than simultaneously so that player may have the advantage of using their peers experience as well.

## 5. RESULT ANALYSIS OF WEEKLY SIMULATION

Simulation results for groups A, B and their respective subgroups (SG1-SG14) are summarized in Tables 2 and 3. Each row in is a representative for each subgroup simulation results showing the total cost summary and order standard deviation pattern for all the four interactive roles (retailer; **R**, wholesaler; W, distributor; D and factory; F). Figure 1(a) shows the simulation results for the wholesaler interactive role at the end of weeks 24 for SG1which is matched with the relevant row in Table 2. Figure 1(b) also depicts order pattern simulation results for SG1and for all components at the end of week 24.

Table- 2. Simulation results for group A & subgroups at the end of week 24

| Group A | Cost summary at end of week 24 | | | | Total Cost (units) | SDT of Order | | | | AVG |
|---|---|---|---|---|---|---|---|---|---|---|
| | R | W | D | F | | R | W | D | F | |
| SG1 | 200 | 241 | 312 | 314 | 1067 | 4.93 | 2.8 | 4.84 | 4.84 | 4.4 |
| SG2 | 246 | 212 | 342 | 285 | 1085 | 4.6 | 3.4 | 5.2 | 5.4 | 4.7 |
| SG3 | 232 | 241 | 222 | 264 | 959 | 4.4 | 3.9 | 3.2 | 4.1 | 3.9 |
| SG4 | 236 | 244 | 301 | 188 | 969 | 4.2 | 3.5 | 3.7 | 4.6 | 4.0 |
| SG5 | 154 | 237 | 324 | 216 | 931 | 3.6 | 2.4 | 2.4 | 3.1 | 2.9 |
| SG6 | 189 | 143 | 232 | 205 | 769 | 2.2 | 1.6 | 1.7 | 2.8 | 2.1 |
| SG7 | 126 | 142 | 155 | 98 | 521 | 2.5 | 2.3 | 1.8 | 1.1 | 1.9 |
| AVG | 197.6 | 208.6 | 269.7 | 224.3 | 6301 | 3.8 | 2.8 | 3.3 | 3.7 | 3.4 |

Table- 3. Simulation results for group B & subgroups at the end of week 24

| Group B | R | W | D | F | Total Cost (units) | SDT of Order | | | | AVG |
|---|---|---|---|---|---|---|---|---|---|---|
| | | | | | | R | W | D | F | |
| SG8 | 75 | 437 | 327 | 382 | 1221 | 5.3 | 5.0 | 5.0 | 5.0 | 5.1 |
| SG9 | 130 | 333 | 286 | 311 | 1060 | 7.2 | 6.2 | 8.0 | 6.9 | 7.1 |
| SG10 | 171 | 195 | 223 | 419 | 1007 | 8.5 | 6.8 | 4.3 | 9.6 | 7.3 |
| SG11 | 377 | 347 | 390 | 243 | 1357 | 10.4 | 7.7 | 6.3 | 5.9 | 7.6 |
| SG12 | 281 | 250 | 275 | 229 | 1035 | 7.7 | 7.0 | 6.8 | 5.6 | 6.8 |
| SG13 | 414 | 359 | 237 | 338 | 1348 | 7.4 | 6.6 | 3.5 | 6.1 | 5.9 |
| SG14 | 268 | 246 | 268 | 312 | 1094 | 8.6 | 5.7 | 8.3 | 9.5 | 8.0 |
| AVG | 245 | 295 | 286 | 334 | 8121 | 7.9 | 6.4 | 6.0 | 6.9 | 6.8 |

The results in Tables 2 and 3 indicate that the value of total cost for group "A" which had the benefit of open internal interaction is 6301 monetary unit while, the respective value for group "B' is 8121 monetary unit. Therefore, the total cost for group A is approximately 30 % lower than group B as shown in Figure 2(b).

The cost and STD trend for group A is more homogenous showing a rather continuous falling down pattern as the simulation progresses from SG1 to SG7. Comparing the cost summary for individual supply chain components also indicates that in the case of group A, the interactive role (highlighted cells) has consistently resulted in reached the minimum cost among other roles while, the same conclusion cannot be made about group B except for SG11.

The average value for each interactive role also reveals that for both groups, the average cost almost progressively increases as one move from downstream (Retailer) to upstream (Factory) which is a good proof for the common bullwhip effect in supply chain practices.

In terms of average standard deviation for orders, it is shown that for group B, the value is twice as much for group A with the values of 3.4 and 6.8 respectively as can be seen in Figure 2(a).

**Figure- 1.** (a) Wholesaler simulation results at the end of week 24 for SG1 and (b) Order pattern simulation for SG1

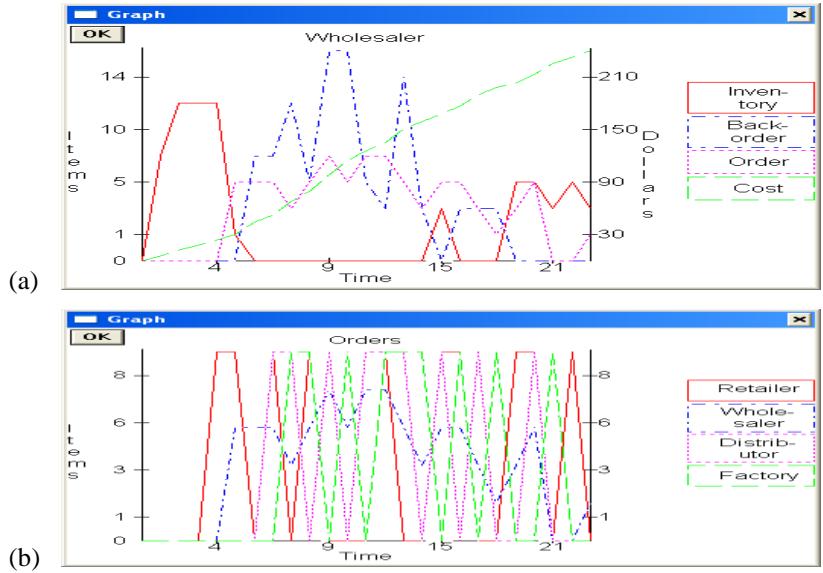

(a)

(b)

**Figure- 2.** (a) Order STD comparison between groups A & B    and    (b) Total cost fluctuation for groups A & B

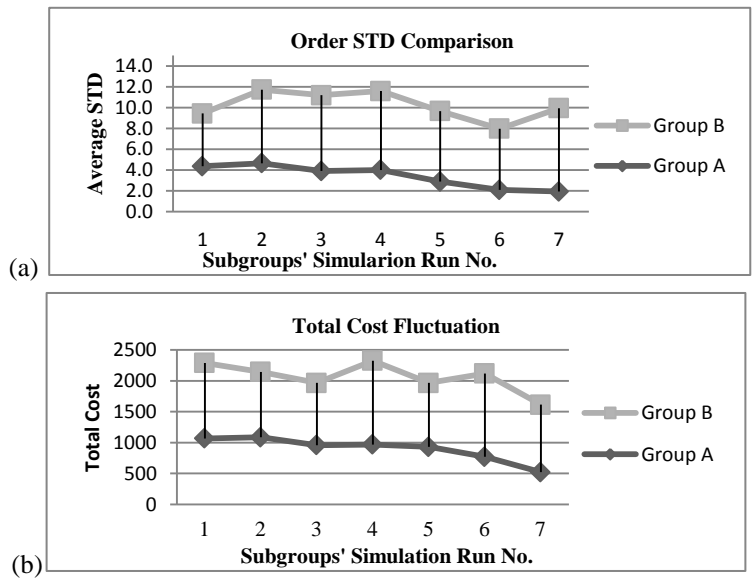

(a)

(b)

## 6. CONCLUSION

In this study the application of beer game simulation software was investigated in supply chain management teaching environment based on a series of structured simulations. Beer game was found to be an effective tool that fully comply with the concept of outcome based education by allowing for independent evaluation of each player's (student's) performance (SCM knowledge). Based on the analysis of cost and order fluctuation reports, it was well proved that mutual interaction among components of a supply chain has significant effects on the overall cost of the chain as well as ordering pattern that is reflected in the corresponding bullwhip effect. The result of the case study shows that having access to upstream and downstream information can play a significant role in reduction of bullwhip effect by minimizing the ordering variation. Lack of such interaction is believed to dramatically intensify supply chain costs as well as the bullwhip effect.

As future work, it is also recommended that beer game be applied in other learning environments and by altering different SCM theories so as to provide a better insight of the real potential of beer game as an OBE-based tool for SCM pedagogy.